# Editorial

## Towards the social media studies of science: social media metrics, present and future

**Rodrigo Costas:** Centre for Science and Technology Studies (CWTS), Leiden University, The Netherlands. rcostas@cwts.leidenuniv.nl

## The rising of new indicators for Science

During the last years a new research topic has rapidly emerged in the field of scientometrics. This new topic, popularly known as *altmetrics,* was first proposed in the *Altmetrics manifesto* (Priem et al., 2010). Since its proposal, altmetrics has been a concept of difficult definition (Haustein, Bowman & Costas, 2016), even being considered as "a good idea, but a bad name" (Rousseau & Ye, 2013). Altmetrics have been usually related to new metrics around scholarly objects captured through events recorded in online social media platforms (Haustein et al., 2016). However, the large diversity of sources and metrics that fall within the realm of altmetrics has made it hard to come up with a consensus of what can be considered as altmetrics (Haustein et al., 2016). *Social media metrics* (SMM) has been seen as one of the best fits as it focuses on the social media perspective from which most of these metrics are captured[1] (Haustein et al., 2016; Wouters, Zahedi & Costas, 2017).

The emergence of SMM has opened a whole new window of possibilities of studying how scientific objects are mentioned, disseminated and discussed in social media. It has even been suggested that they could become a "new discipline" (González-Valiente, Pacheco-Mendoza & Arencibia-Jorge, 2016). In this paper, we aim at providing a general reflection around the present and future of SMM.

A clear indication that research around altmetrics and SMM has boomed during the last few years is the number of reviews around the topic that have been recently published (González-Valiente, Pacheco-Mendoza & Arencibia-Jorge, 2016; Sugimoto et al., 2017; Thelwall & Kousha, 2015; Wouters & Costas, 2012). These reviews have highlighted some of the most critical issues in the development and adoption of SMM. Here we will briefly mention some of them:

- **Sources.** An important body of research has focused on studying the most important sources providing altmetric evidence (Thelwall, Haustein, Larivière, & Sugimoto, 2013; Wouters & Costas, 2012; Zahedi, Costas, & Wouters, 2013). In the last past years several 'altmetric aggregators' such as Altmetric.com (https://www.altmetric.com/), Plum Analytics (http://plumanalytics.com/) or Crossref Event Data (https://www.crossref.org/services/event-data/) have proliferated. These data aggregators focus on the identification and collection of mentions to scholarly objects (mostly scientific publications, books, datasets, etc.) across different social media platforms (e.g., Twitter, Facebook, Mendeley, blogs, or Wikipedia among others).

1. It fails however to incorporate non-social media sources (e.g. newspapers mentions of scientific publications or policy document citations) (Haustein et al., 2016).

- **Coverage.** Another important body of the literature has focused on the study of the coverage of scientific publications across social media platforms (Alperin, 2015; Costas, Zahedi & Wouters, 2015; Haustein, Costas & Larivière, 2015). In general, most results point to a low coverage of scientific publications in social media (e.g. Twitter or Facebook) and relatively higher coverage for more scholarly oriented tools like Mendeley.
- **Correlations and research evaluation possibilities.** Another important issue is the study of the relationship between these new metrics and traditional bibliometric indicators, particularly citations, often in order to discuss the evaluative possibilities of SMM (Costas, Zahedi & Wouters, 2015; Haustein et al., 2014; Thelwall et al., 2013). Most of these studies have shown moderate correlations between Mendeley and citations (Li & Thelwall, 2012; Zahedi et al., 2013; Zahedi, Costas & Wouters, 2017.) and positive but weak correlations for most of the other SMM (e.g. Twitter, Facebook or blogs) (Costas et al., 2015; Haustein et al., 2014). These results support the idea that those sources with a stronger scholarly focus (e.g. Mendeley) could still play some role in supporting or complementing research evaluations; however the evaluative value of the more social media focused indicators, like Twitter or Facebook, it is unclear.
- **Conceptual frameworks.** This weak relationship between most SMM and the more traditional bibliometric indicators has opened the question of what do these SMM actually capture. Haustein et al (2016) provided a first theoretical discussion of SMM in the light of the most common theories considered for citation analysis, showing how the norms that rule scholarly indicators (e.g. citations or peer review) are fundamentally different from those that rule social media behavior. The lack of specific conceptual frameworks around SMM is one of the most important constrains in the development and application of these metrics in real life situations.
- **Other challenges.** Haustein (2016) has highlighted three 'grand challenges' in altmetrics: their *heterogeneity* (reflected in the large diversity of sources, events and metrics that are considered under the umbrella of altmetrics), which hinders the definition of altmetrics and the development of unified conceptual frameworks; the *data quality* issues that challenge the accuracy, comparability and applicability of these metrics; and the *dependencies* on commercial data altmetric aggregators and social media platforms (e.g. Twitter of Facebook, but also ResearchGate or Academica.edu), which make these indicators vulnerable to commercial decisions and the sustainability of these companies. Other challenges surrounding SMM include their easy gaming (e.g.: by automated accounts (Haustein et al., 2015)), or the issues related with their low validity, reliability and transparency (Wouters & Costas, 2012).

## The potential of social media metrics

In spite of these issues, SMM have attracted a lot of attention from many scholarly stakeholders. However, most research so far has depicted a landscape of an unclear utility and validity of social media metrics. It is not only that most SMM have very little relevance in traditional research evaluations (Fraumann, 2017), but also that their potential for more 'societal' evaluation of science (Bornmann, 2013) is still uncertain.

Considering all of the above, it is clear that an important critical question is *what are valid and relevant uses of altmetrics?* To approach this question, recently, more exploratory and descriptive applications of SMM have been discussed (Costas et al., 2017). In these approaches the focus moves from *"how can SMM be used for research evaluation?"* to *"how can SMM inform the reception of science in social media?"*. These more exploratory perspectives open the path towards more strategic uses of altmetric information. Thus, aspects related with the 'who', 'how', 'when' and 'where' of the reception of scientific publications on social media become central. The focus is on monitoring the audiences, reception, perception and discussion of scholarly objects in social media. Examples of descriptive applications include the analysis of communities of attention around scientific publications and topics (Haustein, Bowman, & Costas 2015), *hashtag* analysis (van Honk & Costas, 2016), sentiment analysis (Bae & Lee, 2012), or social media thematic landscapes among others applications (Costas et al., 2017).

These approaches allow, for example, the study of how different Twitter users have different interests on scientific topics. As an example, in figure 1, a map[2] capturing the scientific attention of Twitter users from Spain (below) in contrast with those from Cuba[3] (above) is presented. As it can be seen Cuban tweeters have a stronger interest in papers about economy, management and planning, while Spanish tweeters pay a stronger attention to research about general medicine and sport science, among others.

## The future of social media metrics

When discussing the future of social media metrics there is an important critical challenge that needs to be considered. In line with the notion of dependencies expressed by Haustein (2016), it can also be argued that another form of dependency is linked to the popularity and importance given to social media tools by millions of users around the world. These social media tools are relevant because they are used by large numbers of users. Should (most) Twitter users cease to have any microblogging activity around science, the measurement of the Twitter impact of scientific publications would be inexistent. It is therefore reasonable to argue that the future of SMM is closely tied to the preponderance, scale and importance of social media among users from all over the world. Should these tools lose interest or just being replaced by new tools based on completely different technological approaches; the role, usefulness and value of these SMM will also disappear altogether.

However, the current situation is of an increasing relevance of social media in many different spheres of the scholarly life, with an increasing use of social media tools for scholarly communication purposes and with younger generations of scholars increasingly adopting these new forms of communication (Sugimoto et al., 2017). Thus, scholarly institutions are “increasingly using social media platforms for diffusing and promoting research” (Sugimoto et al., 2017), including among other universities, academic libraries, scientific societies, publishers and individual scholars. It is therefore reasonable to argue that the social media reception (and perception) of scholarly objects is a non-trivial aspect of scientific communication (Wouters et al., 2017). If social media matter, what happens on social media around science, also matters.

From this point of view we can indeed argue that we are witnessing the emergence of a new field. This new field, which could be seen as the *social media studies of science* would be focused on the study of the relationships and interactions between social media and scholarly objects. Thus, research wouldn’t just be circumscribed to the study of the reception of scholarly objects in social media (the predominant approach of most altmetric studies), but also on how scholarly entities interact with other social media actors. In fact, recent developments on the identification of scholars on social media (Costas, van Honk & Franssen, 2017; Ke, Ahn & Sugimoto, 2017) are paving the way to more advanced studies of the interactions between scholarly agents with other social media users. Thus, new potential forms of SMM would include indicators on how scholars are participating in debates in social media, how they engage in the dissemination of scientific information, as well as how scientific organizations are contributing to a better understanding of science through social media tools.

Finally, it is also important to highlight the role that geopolitical factors can play in the access to social media. For example, the limitations and restrictions (being these linguistic, educational, cultural, economic, technical or political) in the access of scholars to social media can contribute to increase the ‘altmetric divide’ (Zahedi, 2017) between richer and poorer countries. Thus, the position of the global North in the scientific debate would be reinforced by a lower awareness, participation and engagement of scholars (as well as citizens) from the less scientifically developed countries in the online social media debate of scientific ideas. ■

2. Map based on the Web of Science Subject Categories. Nodes represent disciplines. Size of the nodes depicts the number of tweets coming from each country. Color indicates disciplines where the presence of tweets from a country is higher than it would be expected by the overall participation of users from the country in the database.

3. It is important to notice that the total amount of tweeters from Cuba that are active discussing scientific publications (as covered in Altmetric.com) is substantially smaller than those from Spain (322 *vs.* 58745). This already can work as an indication of how countries may also face limitations in the access and engagement with publications through social media overall.

**Figure 1. Scholarly interests of Cuban tweeters (above) and Spanish tweeters (below).**

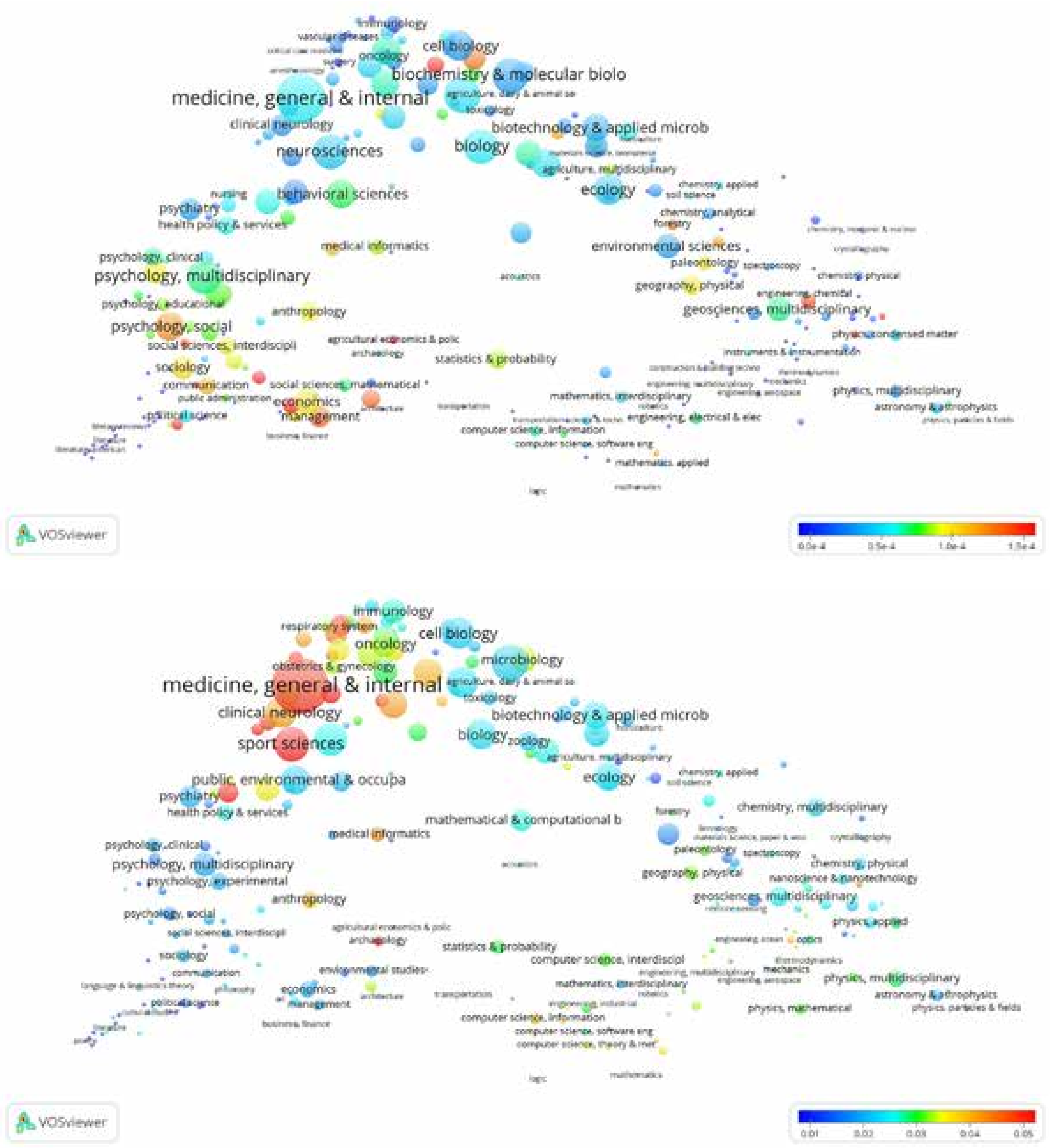

BIBLIOTECA
NACIONAL
DE CUBA
**JOSÉ MARTÍ**

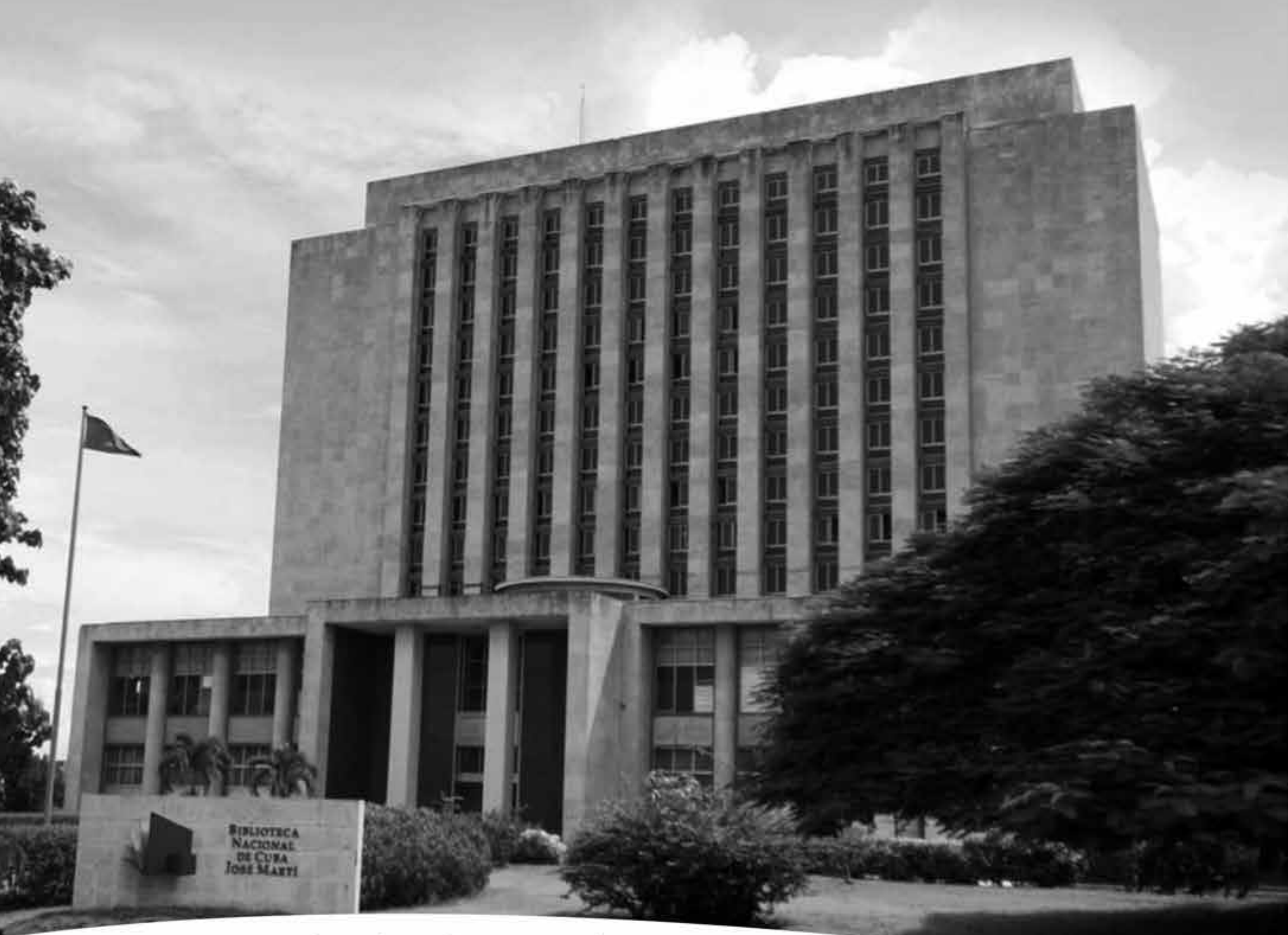

Avenida Independencia y 20 de Mayo,
Plaza de la Revolución, La Habana, Cuba.
Tel.: (53) 7 855 5442 al 49 | dirección@bnjm.cu | www.bnjm.cu
Lunes-Viernes, 8:15 a.m.- 6:15 p.m. / Sábados 8:15 a.m.-4:15 p.m.